\documentclass[aps,preprint,showpacs,nofootinbib,nobibnotes,superscriptaddress]{revtex4}

\usepackage{epsfig}
\usepackage{amssymb}

\begin{document}


\title{Weak and strong regimes of incompressible magnetohydrodynamic turbulence}

\date{\today}

\author{G. Gogoberidze}
\affiliation{Centre for Plasma Astrophysics, K.U.Leuven, Celestijnenlaan
200B, 3001 Leuven, Belgium.}
\affiliation{Georgian National Astrophysical Observatory, 2a Kazbegi
ave., 0160 Tbilisi, Georgia.}

\author{S. M. Mahajan}
\affiliation{Institute for Fusion Studies, University of Texas at Austin, Austin, TX 78712.}

\author{S. Poedts}
\affiliation{Centre for Plasma Astrophysics, K.U.Leuven, Celestijnenlaan
200B, 3001 Leuven, Belgium.}

\begin{abstract}

It is shown that in the framework of the weak turbulence theory, the
autocorrelation and cascade timescales are always of the same order
of magnitude. This means that, contrary to the general belief, any
model of turbulence which implies a large number of collisions among
wave packets for an efficient energy cascade (such as the
Iroshnikov-Kraichnan model) are not compatible with the weak
turbulence theory.

\end{abstract}


\maketitle

\section{Introduction} \label{sec:1}

Magnetohydrodynamic (MHD) turbulence is present in a wide variety of
astrophysical systems such as the solar wind, the interstellar medium,
accretion discs, and so on. Incompressible MHD is the standard
model for the study of astrophysical MHD turbulence. Although
incompressible MHD turbulence has been intensively studied for the
last several decades \cite{I63,Kr65a,SMM83,H84,GS94,GS95,NB97,CV00,GNNP00,MG01,CLV02,MBG03,MG05,OM05,SC05,BL06,G07,BL08}
many physical aspects of the problem still remain unclear.
The first model
of incompressible MHD turbulence was proposed by Iroshnikov
\cite{I63} and Kraichnan \cite{Kr65a}. The Iroshnikov-Kraichnan (IK)
model of MHD turbulence is based on the fact that nonlinear interaction
is possible only among Alfv\'en
waves propagating in opposite directions along the mean magnetic
field. Therefore, an energy cascade occurs as a result of collisions between oppositely
propagating Alfv\'en waves. Consider the isotropic excitation of
Alfv\'en waves on some outer scale $l_0$ with a characteristic
velocity $v_0 \ll V_A$, where $V_A$ denotes the Alfv\'en velocity. The IK
model assumes that the energy transfer is local and isotropic in the
wave number space. The characteristic time scale of the Alfv\'en
wave collision is $\tau_{ac}^{IK} \sim (k V_A)^{-1}$, where $k$ is
the wave number. Using the governing equations of incompressible MHD,
it can be shown that during one collision the distortion of each wave
packet $\delta v_l$ is of the order $\delta v_l/v_l \sim v_l/V_A \ll 1$.
Because these perturbations are summed with random phases $N \sim
(v_l/ \delta v_l)^2 \sim (V_A / v_l)^2$, collisions are necessary to
achieve the distortion of order unity. Therefore, for the energy
cascade time $\tau_{cas}^{IK}$ we have
\begin{equation}
\tau_{cas}^{IK} \sim \frac{1}{k v_l} \frac{V_A}{v_l}. \label{eq:0.2}
\end{equation}
Taking into account the relations $\varepsilon \sim
v_l^2/\tau_{cas}$ and $v_l^2 \sim k \mathcal{E}_k$, where $\varepsilon $ is the
energy cascade rate and $E_k$ is the one dimensional energy
spectrum, we obtain
\begin{equation}
\mathcal{E}_k^{IK} \sim (\varepsilon V_A)^{1/2} k^{-3/2}, \label{eq:0.3}
\end{equation}
which represents the IK spectrum of incompressible MHD turbulence.
Due to the condition $v_0 \ll V_A$ the IK model is usually deemed as
the model of weak MHD turbulence (i.e., the model for which
perturbation theory called weak turbulence theory (WTT)
\cite{K65,ZLF} is applicable).

The IK model for MHD turbulence is isotropic. However, the presence of a
mean magnetic field has a strong effect on the turbulence
properties, in contrast to a mean flow in hydrodynamic turbulence,
which can be eliminated by the corresponding Galilean transformation.
The anisotropy of MHD turbulence had been seen in various
numerical simulations \cite{SMM83,CV00,MG01,MBG03}.
Different theoretical models of weak \cite{GS94,NB97,GNNP00} as well as
strong \cite{H84,GS95,G07,BL08} anisotropic MHD turbulence have
been developed to the date. In Ref.~\cite{SMM83} it was shown that
the lowest order three wave interactions among Alfv\'en waves
are possible only if the parallel (with respect to the constant
magnetic field) wave number of one mode is zero.
This implies that in the framework of WTT there is no parallel cascade
of energy. Consequently, the turbulence is anisotropic and the energy
is cascaded to larger values of the perpendicular wave number.

Comprehensive study of the MHD turbulence in the framework of the
WTT has been performed in Ref. \cite{GNNP00}, where the full coupled
equation for shear and pseudo Alfv\'en waves has been derived  and
analyzed. Stationary solutions of the WTT equations has been found
and was shown that for balanced turbulence the three dimensional
energy spectrum  $E(k_\parallel,k_\perp)\sim
\varepsilon^{1/2}k_\perp^{-3}$. The validity criterium of the WTT
for anisotropic MHD turbulence is \cite{GS95}
\begin{equation}
\frac{k_\perp v_l}{k_\parallel V_A } \ll 1.\label{eq:0.4}
\end{equation}
The usual interpretation of this condition implies that the
nonlinear strain time should be less then the wave period.

In the present paper we study incompressible MHD turbulence in the
framework of the weak coupling approximation (WCA) \cite{K65}, which
represents one of the equivalent formulations of the direct
interaction approximation (DIA) \cite{K59}. The WCA allows us to
study both weak and strong limits of MHD turbulence in the framework
of a unified formalism. In the case of zero residual energy we
rederive the WTT equations from the WCA equations of incompressible
MHD turbulence derived earlier in Ref.~\cite{GNNP00}. We show that
in the framework of the WTT the autocorrelation and cascade
(evolution) timescales are always of the same order of magnitude.
This means that, contrary to the general belief, in the WTT $N \sim
1$ and, consequently, any model for MHD turbulence that implies $N
\gg 1$ (such as the IK model) are incompatible with the WTT.

The paper is organized as follows. The WCA formalism for
incompressible MHD turbulence is described in Sec.~\ref{sec:2}. The
WTT equations are derived and analyzed in Sec.~\ref{sec:3}. The
conclusions are given in Sec.~\ref{sec:5}.

\section{The WCA equations for incompressible MHD turbulence} \label{sec:2}

Consider incompressible MHD turbulence in the presence of a
constant magnetic field ${\bf B}_0$ directed along $z$ axis.
The equations of ideal MHD, governing the evolution of fluctuations of the Elsasser variables,
are
\begin{equation}
\partial_t {\bf U}_1 - {\bf V}_A \partial_z {\bf U}_1 = - ({\bf W}_1 \cdot {\bf \nabla})
{\bf U}_1 - {\bf \nabla}p \label{eq:2.1}
\end{equation}
\begin{equation}
\partial_t {\bf W}_1 + {\bf V}_A \partial_z {\bf W}_1 = - ({\bf U}_1 \cdot {\bf \nabla})
{\bf W}_1- {\bf \nabla}p \label{eq:2.2}
\end{equation}
where ${\bf V}_A \equiv {\bf B_0}/\sqrt{4\pi \rho}$ is the
Alfv\'en velocity, ${\bf W}_1={\bf v}_1+{\bf b}_1$ and ${\bf U}_1={\bf v}_1-{\bf b}_1$  are
fluctuations of the Elsasser variables, ${\bf v}_1$ is the turbulent velocity field, ${\bf
b}_1\equiv {\bf B}_1/\sqrt{4\pi \rho}$ denotes the magnetic field of the fluctuations in velocity
units, $p$ is the pressure
normalized by the density and $\partial_t\equiv \partial/\partial t$.

Performing a Fourier transform, neglecting pseudo Alfv\'en waves
which are known to play only a passive role for strongly anisotropic
MHD turbulence \cite{GS95}, defining the unit polarization vector of
the shear Alfv\'en waves as ${\bf \hat e_{k}}= {\bf \hat k} \times
{\bf z}$, and introducing the amplitudes of the shear Alfv\'en waves
as
\begin{equation}
{\bf w_{\bar k}}=i \phi_{\bf \bar k} {\bf \hat e_{k}},~~~~{\bf
u_{\bar k}}=i \psi_{\bf \bar k} {\bf \hat e_{k}}, \label{eq:5}
\end{equation}
Eqs.~(\ref{eq:2.1})-(\ref{eq:2.2}) reduce to the following set of equations
\begin{equation}
(\omega-\omega_{\bf k}) \phi_{\bf \bar k}= \int_{-\infty}^{\infty}
T_{1,2} \phi_{1} \psi_{2} d{\cal F}_{1,2}^k,
 \label{eq:6.1}
\end{equation}
\begin{equation}
(\omega+\omega_{\bf k}) \psi_{\bf \bar k}= \int_{-\infty}^{\infty}
T_{1,2} \psi_{1} \phi_{2} d{\cal F}_{1,2}^k, \label{eq:6.2}
\end{equation}
where $T_{1,2} \equiv i({\bf \hat e}_{\bf k} \cdot {\bf \hat e}_{\bf
k_1})({\bf k} \cdot {\bf \hat e}_{\bf k_2})$ is the matrix element
of interaction.

Applying the standard technique of WCA, one can then obtain the following set of equations \cite{G07}
\begin{equation}
\zeta_{\bf \bar k}^+ = \int_{-\infty}^\infty |T_{1,2}|^2
 \Gamma_{1}^+ I_2^-
d{\cal F}_{1,2}^k,
 \label{eq:15.1}
\end{equation}
\begin{eqnarray}
-i(\omega-\omega_{\bf k}) I_{\bf \bar k}^+ = {\Gamma_{\bf \bar
k}^+}^\ast \int_{-\infty}^\infty |T_{1,2}|^2 I_1^+ I_2^- d{\cal
F}_{1,2}^k - \nonumber \\ I_{\bf \bar k}^+ \int_{-\infty}^\infty
|T_{1,2}|^2 \Gamma_1^+ I_2^- d{\cal F}_{1,2}^k, \label{eq:15.3}
\end{eqnarray}
and similar equations for $\zeta_{\bf \bar k}^-$ and $I_{\bf \bar
k}^-$. Here ${\bf \bar k} \equiv ({\bf k},\omega)$, the caret
denotes the unit vector, ${\bf u}_1$ denotes ${\bf u_{{\bar
k}_1}}$,~ $\omega_{\bf k}=k_z V_A $ is the frequency of the Alfv\'en
wave, $d {\cal F}_{1,2}^k \equiv d^4 {\bf \bar k}_1 d^4 {\bf \bar
k}_2 \delta_{{\bf \bar k}-{\bf \bar k}_1 - {\bf \bar k}_2}$, and
$\delta_{{\bf \bar k}-{\bf \bar k}_1 - {\bf \bar k}_2} \equiv
\delta({\bf \bar k}-{\bf \bar k}_1 - {\bf \bar k}_2)$ is the Dirac
delta function,
\begin{equation}
\langle \phi_{\bf {\bar k}} \phi_{\bf {\bar k}^\prime}^\ast \rangle
=I_{\bf {\bar k}}^+ \delta_{\bf {\bar k}-\bf {\bar k}^\prime},~~~
\langle \psi_{\bf {\bar k}} \psi_{\bf {\bar k}^\prime}^\ast \rangle
=I_{\bf {\bar k}}^- \delta_{\bf {\bar k}-\bf {\bar k}^\prime},
 \label{eq:7}
\end{equation}
and
\begin{equation}
\Gamma_{\bf \bar k}^\pm =  \frac{i}{\omega \mp \omega_{\bf k} +
i\zeta_{\bf \bar k}^\pm}.
 \label{eq:16}
\end{equation}

Eqs.~(\ref{eq:15.1})-(\ref{eq:15.3}) are useless unless some
assumptions are made about the frequency dependence of $\Gamma_{\bf
\bar k}^\pm$ and $I_{\bf \bar k}^\pm$. Equivalently, in the
framework of the DIA one should make some assumptions about the time
dependence of $G^\pm({\bf k},\tau)$ and $Q^\pm({\bf k},\tau)$
\cite{Le}, which are the corresponding inverse Fourier transforms of
$\Gamma_{\bf \bar k}^\pm/2\pi$ and $I_{\bf \bar k}^\pm$ with respect
to $\omega$ [i.e., $\langle \phi_{\bf k}(t+\tau) \phi_{\bf
k^\prime}^\ast(t) \rangle =Q^+({\bf k},\tau) \delta_{{\bf k}-{\bf
k}^\prime}$] . One of the simplest and frequently used assumptions
implies \cite{K59,Le}
\begin{equation}
G^\pm({\bf k},\tau)=\exp{\left( -|\eta_{\bf k}^\pm| \tau \pm i
\omega_{\bf k} \tau \right)} H(\tau), \label{eq:36.1}
\end{equation}
\begin{equation}
Q^\pm({\bf k},\tau)=\exp{\left(-|\eta_{\bf k}^\pm|\tau \pm i
\omega_{\bf k} \tau \right)} E_{\bf k}, \label{eq:36.2}
\end{equation}
where $H(t)$ is the Heaviside (step) function, and $E_{\bf
k}$ is the energy spectrum.

Similar to Eqs.~(\ref{eq:36.1})-(\ref{eq:36.2}), in the case under
consideration we assume
\begin{equation}
\zeta_{\bf \bar k}^\pm=\eta_{\bf k}^\pm, \label{eq:37.1}
\end{equation}
\begin{equation}
I_{\bf \bar k}^\pm = \frac{E_{\bf k}^\pm}{\pi}
\frac{\eta_{\bf k}^\pm}{(\omega \mp \omega_{\bf k})^2 + ({\eta_{\bf
k}^\pm})^2}. \label{eq:37.2}
\end{equation}
Here, for simplicity, we consider the symmetric case $\eta_{\bf k}^\pm=\eta_{\bf k}$,
and $E_{\bf k}^\pm=E_{\bf k}$, which physically corresponds
to a turbulence with zero cross helicity. Although we consider the symmetric
case, in the further analysis we will keep the $\pm$ signs for the energy spectra in order
to underline the fact that nonlinear interactions are possible only between
counter propagating modes.

$\tau_{ac} \equiv 1/\eta_{\bf k}$ is the autocorrelation time. As it
was mentioned in Refs.~\cite{K65,Kr65a} the random Galilean
invariance requires that before applying the WCA (DIA) closure
scheme to Eqs.~(\ref{eq:6.1})-(\ref{eq:6.2}) one should remove the
influence of the velocity field of low frequency modes (for a more
detailed analysis see Ref.~\cite{G07}). If this is done, and the
corresponding contributions are removed from
Eqs.~(\ref{eq:15.1})-(\ref{eq:15.3}), then $\tau_{ac}$ represents
the Lagrangian autocorrelation timescale, which is called "the
duration of unit act of interaction" in heuristic models of the
turbulence.

\section{Derivation and analysis of the WTT equations} \label{sec:3}

Substituting Eqs.~(\ref{eq:37.1}) and (\ref{eq:37.2}) into Eq.~(\ref{eq:15.1}) and performing
an integration with respect to the frequencies, we get
\begin{eqnarray}
-i(\omega-\omega_{\bf k}) I_{\bf \bar k}^+ = {\Gamma_{\bf \bar
k}^+}^\ast \int_{-\infty}^\infty |T_{1,2}|^2 R_a E_1^+ E_2^- {\rm d}\mathcal{K}_{1,2}^k -
\nonumber \\ I_{\bf \bar k}^+ \int_{-\infty}^\infty
|T_{1,2}|^2 R_b E_2^- {\rm d}\mathcal{K}_{1,2}^k, \label{eq:s2.1}
\end{eqnarray}
\begin{equation}
\eta_{\bf k} = \int_{-\infty}^\infty
|T_{1,2}|^2 R_b E_2^- {\rm d}\mathcal{K}_{1,2}^k,
 \label{eq:s2.2a}
\end{equation}
where
\begin{equation}
R_a = \frac{1}{\pi} \frac{\eta_1+\eta_2}{(\eta_1+\eta_2)^2 + (\omega-\omega_k+ \Delta \omega_{k12})^2},
 \label{eq:s2.2}
\end{equation}
\begin{equation}
R_b = \frac{1}{(\eta_1+\eta_2)- i(\omega-\omega_k+ \Delta \omega_{k12})},
 \label{eq:s2.3}
\end{equation}
${\rm d} \mathcal{K}_{1,2}^k \equiv {\rm d}^3{\bf k}_1 {\rm
d}^3{\bf k}_2 \delta_{{\bf k}- {\bf k}_1 -{\bf k}_2}$, and $\Delta \omega_{k12}=\omega_{\bf k}-\omega_{{\bf k}_1}+\omega_{{\bf k}_2}$.

According to Eq.~(\ref{eq:37.2}), $I_{\bf \bar k}^\pm$ significantly
differs from zero when $(\omega-\omega_{\bf k}) \lesssim \eta_{\bf
k}$. First of all, consider the real part of the right hand side of
Eq.~(\ref{eq:s2.1}) which describes the nonlinear decay of the
fluctuations (whereas the real part describes the frequency shift
caused by nonlinear interactions). Consider the limit $\eta_{\bf k}
\rightarrow 0$. In this case, the width of the wave packets tends to
zero and Eq.~(\ref{eq:37.2}) yields $I_{\bf \bar k}^\pm=E^\pm_{\bf
k} \delta_{\omega \mp \omega_{\bf k}}$. In the considered limit the
integrals on the right hand side of Eq.~(\ref{eq:s2.1}) are
dominated by the contribution of the small vicinity of the so-called
resonant curve (defined by the condition $\Delta \omega_{k12}=0$)
where the condition
\begin{equation}
\Delta \omega_{k12} \lesssim \eta_{\bf k},
 \label{eq:s2.4}
\end{equation}
is fulfilled. The solution of the resonant condition $\Delta \omega_{k12}=0$
is $k_{z2}=0$. Consequently, three wave resonant interactions must include
the zero frequency mode \cite{SMM83,NB97,GNNP00}. The volume of the wave number space occupied
by the resonant area where the condition (\ref{eq:s2.4}) is fulfilled
is $[\eta_{\bf k}/(\partial \omega_{k}/\partial k_z)] k_\perp^2$.
Taking also into account that a typical value of $R_a$ and $\Re{(R_b)}/\pi$ in the resonant area is $1/\eta_{\bf k}$, noting that $T_{1,2} \sim k_\perp$ and assuming that
the nonlinear energy transfer in the $ {\bf k}_\perp$ plane is dominated
by triad interactions with $k_{\perp} \sim k_{\perp1,2}$
then the contribution of the resonant area in (say) the first integral
of Eq.~(\ref{eq:s2.1}) can be estimated as $k_\perp^4 E^+(k_z,k_\perp)
E^-(0,k_\perp) /(\partial \omega_{k}/\partial k_z)$. Similarly,
the contribution of the rest part of the ${\bf k}$ space is $\eta_{\bf k} k_z
k_\perp^4 E^+(k_z,k_\perp) E^-(k_z,k_\perp)
/\omega_{k}^2$. Consequently, the domination of the resonant contribution implies
\begin{equation}
\frac{\eta_{\bf k}}{\omega_{\bf k}} \frac{E^-(k_z,k_\perp)}{E^-(0,k_\perp)} \ll 1.
 \label{eq:s2.4a}
\end{equation}
If this condition is fulfilled, one can replace $R_a$ and $\Re{(R_b)}/\pi$ by
$\delta(\omega_{\bf k}-\omega_{{\bf k}_1}+\omega_{{\bf k}_2}) \equiv \delta_{\Delta \omega_{k12}}$.
Then, the real part of Eq.~(\ref{eq:s2.1}) reduces to
\begin{equation}
-\gamma_{\bf k} E_{\bf k}^+ = \pi \int_{-\infty}^\infty |T_{1,2}|^2
E_2^-({E}_1^+  - {E}_{\bf k}^+)
\delta_{\Delta \omega_{k12}}{\rm d}\mathcal{K}_{1,2}^k, \label{eq:s2.5}
\end{equation}
where $\gamma_{\bf k}$ is the total decrement caused by the nonlinear interactions.

In contrast to the DIA which operates with a two point two time
correlation functions and/or their Fourier transforms ($Q^\pm({\bf
k},\tau)$ and $I_{\bf \bar k}^\pm$), the WTT implies Markovian
closure and consequently operates with the Fourier transform of a
two point one time correlation function $n^\pm_{\bf k}(t)$ defined
as $\langle \phi_{\bf k}(t) \phi_{\bf k^\prime}^\ast(t) \rangle
=n^\pm_{\bf k}(t) \delta_{{\bf k}-{\bf k}^\prime}$. As known, the
nonlinear decrement of a one time correlation function for a zero
time separation is twice larger then the nonlinear decrement of a
two point correlation function since all temporal derivatives in the
dynamic equations now act on both the time variables \cite{Le}.
Consequently, according to Eq.~(\ref{eq:s2.5}) the dynamic equation
for $n^+_{\bf k}(t)$ is
\begin{equation}
\frac{\partial n^+_{\bf k}}{\partial t} = 2\pi \int_{-\infty}^\infty |T_{1,2}|^2
n^-_2 (n^+_1  - n^+_{\bf k})
\delta_{\Delta \omega_{k12}}{\rm d}\mathcal{K}_{1,2}^k. \label{eq:s2.7}
\end{equation}
Similar manipulations lead to the following equation for $n^-_{\bf k}(t)$
\begin{equation}
\frac{\partial n^-_{\bf k}}{\partial t} = 2\pi \int_{-\infty}^\infty |T_{1,2}|^2
n^+_2 (n^-_1  - n^-_{\bf k})
\delta_{\Delta \omega_{k12}}{\rm d}\mathcal{K}_{1,2}^k. \label{eq:s2.8}
\end{equation}
These equations represent the WTT equations for weak MHD turbulence (with zero
residual energy), which was first derived in Ref.~\cite{GNNP00}
using the standard WTT technique.

Let us now turn back to the imaginary part of Eqs.~(\ref{eq:s2.1}).
For validity of the WTT it is necessary that the frequency shift
caused by nonlinear interactions is smaller than $\omega_{\bf k}$.
Taking into account that, according to Eq.~(\ref{eq:s2.2a}),
$\eta_{\bf k} \sim  k_\perp^4 {E}^-(0,k_\perp)/\partial \omega_{\bf
k}/\partial k_z$,  an analysis similar to the one performed above
shows that, in addition to Eq.~(\ref{eq:s2.4a}), the following
condition should be satisfied
\begin{equation}
\frac{\eta_{\bf k}}{\omega_{\bf k}} \ll 1.
\label{eq:s210}
\end{equation}

The validity conditions of the WTT in a form similar to
Eqs.~(\ref{eq:s2.4a}) and (\ref{eq:s210}) were first derived in
Ref.~\cite{K65}. In general case, the validity criterium of the WTT
was found to  be identical to Eq.~(\ref{eq:s210}). The appearance of
the additional condition (\ref{eq:s2.4a}) for the validity of the
kinetic equation (\ref{eq:s2.7}), is related to the degenerate
character of the solution of the resonant condition $\omega_{\bf
k}=\omega_{{\bf k}_1}-\omega_{{\bf k}_2}$, which implies $k_{2z}=0$,
and therefore requires the participation in nonlinear interactions
of modes with very low frequencies [in the context of the performed
analysis it is clear that ${E}^-(0,k_\perp)$ in Eq.~(\ref{eq:s2.4a})
should be understood as the average energy density of the modes with
$k_z \lesssim \eta_{\bf k}/(\partial \omega_{\bf k}/\partial k_z)$].
If Eq.~(\ref{eq:s210}) holds but the intensity of the low frequency
modes is very low, such that left hand side of Eq.~(\ref{eq:s2.4a})
is much greater then unity, then the kinetic equation
(\ref{eq:s2.7}) is not valid. However, it can be shown that the WTT
is still valid then and the nonlinear interactions are dominated by
four wave interactions. The corresponding kinetic equation was
derived in Ref.~\cite{GS94}. Introducing the characteristic velocity
of the perturbations with a characteristic parallel length scale
$l_z \sim 1/k_z$ and perpendicular length scale $l_\perp \sim
1/k_\perp$, respectively, as $v_l^2 \sim k_z k_\perp^2
{E}(k_z,k_\perp)$, the conditions (\ref{eq:s2.4a}) and
(\ref{eq:s210}) reduce to the following ones $k_\perp v_l/k_zV_A \ll
1$ and $[k_\perp v_l E(0,k_\perp)]/[k_zV_A E(k_z,k_\perp)]
 \ll 1$. Note that the first condition coincides with Eq.~(\ref{eq:0.4}).

Another important temporal characteristic of the turbulence
(together with $\tau_{ac}$) is the energy cascade timescale
$\tau_{cas}$ which represents the timescale of the energy cascade
described by Eq.~(\ref{eq:s2.7}). A simple way of determining
$\tau_{cas}$ is the following (a mathematically more sound
derivation can be found in Ref.~\cite{ZLF}): in the case of
stationary turbulence, the left hand side of Eq.~(\ref{eq:s2.7}) is
zero. But the characteristic timescale of the energy cascade can be
determined if we retain only the modes with $|{\bf k}_{1\perp}|
>|{\bf k}_{\perp}|$ on the right hand side of Eq.~(\ref{eq:s2.7}).
The equation then obtained describes the energy transfer from the
mode with a wave number ${\bf k}_\perp$ to the modes with higher
perpendicular wave numbers. The right hand side can be estimated as
$n^+/\tau_{cas}$. The analysis of the left hand side terms similar
to the one performed for the estimation of $\eta_{\bf k}$ yields
\begin{equation}
\tau_{cas} \sim \tau_{ac},
 \label{eq:s2.11}
\end{equation}
and consequently, the WTT always suggests $N \sim (\tau_{cas} /
\tau_{ac})^2 \sim 1$. These arguments show that the IK model, which
implies $\tau_{ac} \sim (k V_A)^{-1}$, $\tau_{cas} \sim V_A/(k
v_l^2)$ and consequently $N \sim (V_A/v_l)^2 \gg 1$, are
incompatible with the WTT. Note that in the considered isotropic
case Eq.~(\ref{eq:0.4}) leads to the (incorrect) conclusion that the
turbulence is weak. On the other hand, because $\tau_{ac}k V_A \sim
1$ Eq.~(\ref{eq:s210}) yields that the IK model does not correspond
to the weak turbulence limit.

As was shown above, the width of the resonant are is $\Delta k_z \sim
(\eta_{\bf k}/\omega_{\bf k})k_z$. Taking into account Eq.~(\ref{eq:s210}) this implies that physically nonlinear interactions
in the framework of the WTT can be interpreted as resonant interactions
among spatially very large wave packets [with characteristic size
$\Delta l \sim (\omega_{\bf k}/\eta_{\bf k})/k_z$]. Although the
nonlinear interactions are weak, the interacting wave packets are very
large such that an original wave packet decays before interacting
wave packets pass through each other. Because the introduction of the Dirac
delta function in Eq.~(\ref{eq:s2.7}) requires the limit $\eta_{\bf
k}/\omega_{\bf k} \rightarrow 0$, it is sometimes stated that in the
framework of the WTT the units of the nonlinear interactions are not wave
packets of a finite spatial extent but spatially infinite Fourier
harmonics \cite{K65}.

\section{Conclusions} \label{sec:5}

Both strong and weak turbulence regimes of incompressible MHD
turbulence were considered in the framework of the WCA. We showed
that in the framework of the WTT the autocorrelation and cascade
timescales are always of the same order of magnitude and,
consequently, the WTT always suggests $N \gg 1$. Physically this is
caused by the fact that the framework of the WTT can be interpreted
as the resonant interaction among spatially very large wave packets
[with characteristic size $\Delta l \sim (\omega_{\bf k}/\eta_{\bf
k})/k_z$]. Although the nonlinear interactions are weak, the
interacting wave packets are very large such that an original wave
packet decays before interacting wave packets pass through each
other.

\begin{acknowledgments}
G.G. and S.M.M. acknowledge the hospitality of the Abdus Salam International
Center for Theoretical Physics (ICTP) where part of the work was done.
G.G. acknowledges partial support from INTAS grant 061000017-9258
and Georgian NSF grants ST06/4-096 and ST07/4-193.
S.M.M.'s work was supported by USDOE Contract No.~DE--FG02--04ER--54742.
These results were obtained in the framework of the projects
GOA/2009-009 (K.U.Leuven), G.0304.07 (FWO-Vlaanderen) and
C~90205 (ESA Prodex 9).
Financial support by the European Commission through the SOLAIRE
Network (MTRN-CT-2006-035484) is gratefully acknowledged.

\end{acknowledgments}



\end{document}